# Multi-party quantum private comparison based on entanglement swapping of Bell entangled states within $d$-level quantum system

Tian-Yu Ye*, Jia-Li Hu

College of Information & Electronic Engineering, Zhejiang Gongshang University, Hangzhou 310018, P.R.China

**Abstract:** In this paper, a multi-party quantum private comparison (MQPC) scheme is suggested based on entanglement swapping of Bell entangled states within $d$-level quantum system, which can accomplish the equality comparison of secret binary sequences from $n$ users via one execution of scheme. Detailed security analysis shows that both the outside attack and the participant attack are ineffective. The suggested scheme needn't establish a private key among $n$ users beforehand through the quantum key distribution (QKD) method to encrypt the secret binary sequences. Compared with previous MQPC scheme based on $d$-level Cat states and $d$-level Bell entangled states, the suggested scheme has distinct advantages on quantum resource, quantum measurement of third party (TP) and qubit efficiency.

## 1 Introduction

In the year of 1982, Yao[1] put forward the millionaires' problem, i.e., two millionaires want to know who is richer without leaking out their actual properties. Afterward, Boudot *et al.* [2] designed a scheme to judge whether two millionaires are equally rich or not. The problems both Yao[1] and Boudot *et al.*[2] focused on belong to classical private comparison. In the year of 2009, Yang *et al.* [3] put forward the concept of quantum private comparison (QPC) for the first time by combining quantum mechanics and classical private comparison. From then on, QPC has entered into the eyes of researchers so that lots of two-party QPC schemes[4-13] have been gradually designed.

However, the two-party QPC scheme always has a drawback, i.e., if it is used to accomplish the equality comparison of private inputs from $n$ parties, it has to be implemented for $(n-1) \sim n(n-1)/2$ times. In order to accomplish this task within one execution of protocol, Chang *et al.*[14] constructed the first multi-party quantum private comparison (MQPC) scheme by using $n$-particle GHZ class states in the year of 2013. From then on, the MQPC which accomplishes the equality comparison of different secrets has gained rapid developments[15-21]. In the year of 2014, Liu *et al.* [15] suggested a MQPC scheme based on $d$-level $n$-particle entangled states; Wang *et al.*[16] designed two MQPC schemes based on $n$-level entangled states. In the year of 2017, Hung *et al.*[17] constructed a MQPC scheme for strangers with almost dishonest third parties (TP). We successively put forward the MQPC schemes based on the entanglement swapping of Bell entangled states[18], the entanglement swapping between $d$-level Cat states and $d$-level Bell states[19], scattered preparation and one-way convergent transmission of quantum states[20], and $n$-level single-particle states[21].

Based on the above analysis, this paper concentrates on constructing a MQPC scheme based on entanglement swapping of Bell entangled states within $d$-level quantum system, which can also accomplish the equality comparison of secret binary sequences from $n$ users via one execution of scheme. The proposed scheme takes great advantages over the scheme of Ref.[19] on the aspects of quantum resource, quantum measurement of TP and qubit

---

*Corresponding anthor：
 E-mail：happyyty@aliyun.com

efficiency.

## 2 The entanglement swapping of Bell entangled states within $d$-level quantum system

Bell entangled state within $d$-level quantum system is defined as

$$|\phi(u,v)\rangle = \frac{1}{\sqrt{d}} \sum_{j=0}^{d-1} \zeta^{ju} |j, j\oplus v\rangle, \tag{1}$$

where $u,v \in \{0,1,\ldots,d-1\}$, $\oplus$ is the modulo $d$ addition and $\zeta = e^{2\pi i/d}$. $|\phi(u,v)\rangle$ can be generated by performing $U_{(u,v)}$ on $|\phi(0,0)\rangle$. i.e.,

$$(I \otimes U_{(u,v)})|\phi(0,0)\rangle = |\phi(u,v)\rangle, \tag{2}$$

where

$$U_{(u,v)} = \sum_{j=0}^{d-1} \zeta^{ju} |j\oplus v\rangle\langle j|. \tag{3}$$

The entanglement swapping of two Bell entangled states within $d$-level quantum system can be represented by

$$|\phi(u_1,u_2)\rangle_{12} |\phi(v_1,v_2)\rangle_{34} = \frac{1}{d} \sum_{k,l=0}^{d-1} \zeta^{kl} |\phi(u_1\oplus k, v_2\oplus l)\rangle_{14} |\phi(v_1\oplus(-k), u_2\oplus(-l))\rangle_{32}, \tag{4}$$

just as shown in Fig 1, where each solid circle or each hollow circle denotes one particle, and two particles connected by one solid line form one Bell entangled state within $d$-level quantum system.

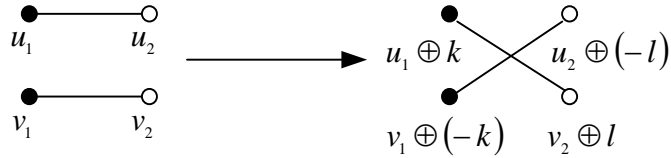

Fig 1  The entanglement swapping of two Bell entangled states within $d$-level quantum system

## 3 The proposed scheme
### 3.1 Scheme description

Suppose that the $i^{\text{th}}$ user, $P_i$, has a secret binary sequence $K_i$ of length $N$, i.e., $K_i = (k_i^1, k_i^2, \ldots, k_i^N)$, where $k_i^t \in \{0,1\}$, $i = 1,2,\ldots,n$ and $t = 1,2,\ldots,N$. $n$ users want to accomplish the equality comparison of their secret binary sequences with the help of a semi-honest TP, $P_0$. Here, the term 'semi-honest' means that TP tries her best to obtain $K_i$ when implementing the scheme but cannot be allowed to conspire with others[8]. In the proposed scheme, it is assumed that $d = n$.

**Step 1:** $P_0$ and $P_1$ **mutually transmit particle sequences and swap entanglement**

(1) $P_0$ and $P_1$ mutually transmit particle sequences

$P_0$ generates $N$ $d$-level Bell entangled states $|\phi(0,0)\rangle$. $P_0$ picks out all of the first and the second particles of these Bell entangled states to form two ordered sequences, $S_0^1 = \{u_0^1, u_0^2, \ldots, u_0^N\}$ and $S_0^2 = \{v_0^1, v_0^2, \ldots, v_0^N\}$, respectively. $P_0$ generates one group of decoy photons[22,23] according to $V_1 = \{|r\rangle\}_{r=0}^{d-1}$ and $V_2 = \{F|r\rangle\}_{r=0}^{d-1}$, where $r \in \{0,1,\ldots,d-1\}$ and $F$ is the $d^{\text{th}}$ order discrete quantum Fourier transform, and randomly inserts them into $S_0^2$ to form a new sequence $S_0^{2'}$. Finally, $P_0$ transmits $S_0^{2'}$ to $P_1$.

$P_1$ generates $N$ $d$-level Bell entangled states $|\phi(0,0)\rangle$. Then, in order to encode $k_1^t$, $P_1$ imposes $I \otimes U_{(0,k_1^t)}$ on the $t^{th}$ ($t=1,2,\ldots,N$) Bell entangled states to produce $|\phi(0,k_1^t)\rangle$. $P_1$ picks out all of the first and the second particles of these encoded Bell entangled states to form two ordered sequences, $S_1^1 = \{u_1^1, u_1^2, \ldots, u_1^N\}$ and $S_1^2 = \{v_1^1, v_1^2, \ldots, v_1^N\}$, respectively. $P_1$ generates one group of decoy photons according to $V_1$ and $V_2$, and randomly inserts them into $S_1^2$ to form a new sequence $S_1^{2'}$. Finally, $P_1$ transmits $S_1^{2'}$ to $P_0$.

(2) $P_0$ and $P_1$ implement eavesdropping check

$P_0$ and $P_1$ utilize the corresponding decoy photons to check the transmission security of $S_0^{2'}$ and $S_1^{2'}$, respectively.

After $P_1$ receives $S_0^{2'}$, $P_1$ and $P_0$ check the transmission security of $S_0^{2'}$ together. $P_0$ tells $P_1$ the positions and the preparation basis of decoy photons in $S_0^{2'}$. $P_1$ uses the preparation basis of $P_0$ to measure the decoy photons in $S_0^{2'}$ and tells $P_0$ their measurement results. Then, $P_0$ judges whether an eavesdropper is on line during the transmission of $S_0^{2'}$ by comparing the prepared initial states of decoy photons in $S_0^{2'}$ with their measurement results of $P_1$. If there is no eavesdropper, they will continue the communication; otherwise, the communication will be terminated.

After $P_0$ receives $S_1^{2'}$, $P_0$ and $P_1$ also use the above eavesdropping check method to check the transmission security of $S_1^{2'}$ together. If there is no eavesdropper, they will continue the communication; otherwise, the communication will be terminated.

(3) $P_0$ and $P_1$ swap entanglement for particle sequences

$P_0$ discards the decoy photons in $S_1^{2'}$ to restore $S_1^2$. $P_1$ discards the decoy photons in $S_0^{2'}$ to restore $S_0^2$. $P_1$ performs $d$-level Bell entangled state measurement on particles $u_1^t$ and $v_0^t$ to obtain the value of $-l_1^t$.

**Step 2:** $P_0$ **and** $P_j$ ($j=2,3,\ldots,n$) **mutually transmit particle sequences and swap entanglement**

(1) $P_0$ and $P_j$ mutually transmit particle sequences

$P_j$ generates $N$ $d$-level Bell entangled states $|\phi(0,0)\rangle$. Then, in order to encode $k_j^t$, $P_j$ imposes $I \otimes U_{(0,k_j^t)}$ on the $t^{th}$ ($t=1,2,\ldots,N$) Bell entangled states to produce $|\phi(0,k_j^t)\rangle$. $P_j$ picks out all of the first and the second particles of these encoded Bell entangled states to form two ordered sequences, $S_j^1 = \{u_j^1, u_j^2, \ldots, u_j^N\}$ and $S_j^2 = \{v_j^1, v_j^2, \ldots, v_j^N\}$, respectively. $P_j$ generates one group of decoy photons according to $V_1$ and $V_2$, and randomly inserts them into $S_j^2$ to form a new sequence $S_j^{2'}$. Finally, $P_j$ transmits $S_j^{2'}$ to $P_0$.

$P_0$ generates one group of decoy photons according to $V_1$ and $V_2$, and randomly inserts them into $S_{j-1}^2$ to form a new sequence $S_{j-1}^{2'}$. Finally, $P_0$ transmits $S_{j-1}^{2'}$ to $P_j$.

(2) $P_0$ and $P_j$ implement eavesdropping check

$P_0$ and $P_j$ utilize the corresponding decoy photons to check the transmission security of $S_j^{2'}$ and $S_{j-1}^{2'}$, respectively.

After $P_0$ receives $S_j^{2'}$, $P_0$ and $P_j$ uses the eavesdropping check method same to that of Step 1 to check the transmission security of $S_j^{2'}$. If there is no eavesdropper, they will continue the communication; otherwise, the communication will be terminated.

After $P_j$ receives $S_{j-1}^{2"}$, $P_j$ and $P_0$ uses the eavesdropping check method same to that of Step 1 to check the transmission security of $S_{j-1}^{2"}$. If there is no eavesdropper, they will continue the communication; otherwise, the communication will be terminated.

(3) $P_0$ and $P_j$ swap entanglement for particle sequences

$P_0$ discards the decoy photons in $S_j^{2'}$ to restore $S_j^2$. $P_j$ discards the decoy photons in $S_{j-1}^{2"}$ to restore $S_{j-1}^2$. $P_j$ performs $d$-level Bell entangled state measurement on particles $u_j^t$ and $v_{j-1}^t$ to obtain the value of $k_{j-1}^t \oplus l_{j-1}^t \oplus \left(-l_j^t\right)$. $P_0$ performs $d$-level Bell entangled state measurement on particles $u_0^t$ and $v_n^t$ to obtain the value of $k_n^t \oplus l_n^t$.

**Step 3: Equality comparsion**

$P_0, P_1, \ldots, P_n$ privately cooperate to compute

$$sum_t' = \left(-l_1^t\right) \oplus \left(k_1^t \oplus l_1^t \oplus \left(-l_2^t\right)\right) \oplus \left(k_2^t \oplus l_2^t \oplus \left(-l_3^t\right)\right) \oplus \ldots \oplus \left(k_{n-1}^t \oplus l_{n-1}^t \oplus \left(-l_n^t\right)\right) = k_1^t \oplus k_2^t \oplus \ldots \oplus k_{n-1}^t \oplus \left(-l_n^t\right), \quad (5)$$

and informs $P_0$ of the value of $sum_t'$ via a public channel. Then, $P_0$ calculates Eq.(6) and obtains the value of $k_1^t \oplus k_2^t \oplus \ldots \oplus k_n^t$ ($t = 1, 2, \ldots, N$):

$$sum_t = k_1^t \oplus k_2^t \oplus \ldots \oplus k_{n-1}^t \oplus \left(-l_n^t\right) \oplus \left(k_n^t \oplus l_n^t\right) = k_1^t \oplus k_2^t \oplus \ldots \oplus k_{n-1}^t \oplus k_n^t. \quad (6)$$

If $sum_t = 0$, it will have $k_1^t = k_2^t = \ldots = k_n^t = 0$ or $k_1^t = k_2^t = \ldots = k_n^t = 1$, which means that $k_1^t, k_2^t, \ldots, k_n^t$ are all equal; otherwise, $k_1^t, k_2^t, \ldots, k_n^t$ are not all equal. If $sum_t = 0$ for $t = 1, 2, \ldots, N$, $P_0$ will announce that $K_1, K_2, \ldots, K_n$ are all equal; otherwise, $P_0$ will announce that $K_1, K_2, \ldots, K_n$ are not all equal.

## 3.2 Security analysis

### (1) Outside attack

In the proposed scheme, $P_0$ and $P_i$ ($i = 1, 2, \ldots, n$) mutually transmit quantum state sequences, and use the $d$-level decoy photons randomly chosen from $V_1$ and $V_2$ to detect the existence of an outside eavesdropper. The decoy photon technology[22,23] has been extensively used to guarantee the security of quantum cryptography scheme. It can be regarded as a variant of the eavesdropping check method of BB84 quantum key distribution (QKD) scheme[24], which has been proven to have unconditional security[25]. Thus, the proposed scheme is immune to the active attacks from an outside eavesdropper.

In Step 3, $P_0$ announces the comparison result of $K_1, K_2, \ldots, K_n$. Fortunately, an outside eavesdropper cannot deduce $K_i$ just from the comparison result of $K_1, K_2, \ldots, K_n$.

### (2) Participant attack

The attack from a dishonest participant is always more powerful, since she joins in the implementation of scheme. Thus, this kind of attack should be paid more attention to[26]. In the following, three kinds of participant attack are analyzed in detail.

① The attack from one dishonest user

In the proposed scheme, each user takes the following actions: generating quantum state sequence and encoding her binary sequence on it, mutually transmitting quantum state sequence and implementing eavesdropping check processes together with $P_0$, swapping entanglement for quantum state sequences without decoy photons together with $P_0$, and cooperating with other users to calculate Eq.(5) privately. It is easy to find out that, the role of each user is similar. Suppose that $P_i$ ( $i \in \{1,2,\ldots,n\}$ )is a dishonest user. $P_i$ may launch her active attacks on particles of $S_v^{2'}$ ( $v = 1,2,\ldots,n$ and $v \neq i$ ) and $S_u^{2''}$ ( $u = 1,2,\ldots,n-1$ and $u \neq i$ ). In this case, $P_i$ actually acts as an outside eavesdropper. As a result, her attacks inevitably leave trace on decoy photons so that they are detected by the eavesdropping check processes undoubtedly.

On the other hand, in Step 3, $P_0$ announces the comparison result of $K_1, K_2, \ldots, K_n$. However, $P_i$ cannot deduce $K_v$ just from the comparison result of $K_1, K_2, \ldots, K_n$.

② The collusion attack from two or more dishonest users

Here, we consider the most extreme situation, i.e., $n-1$ users collude to obtain the secret binary sequence of the left user. Without loss of generality, assume that $n-1$ dishonest users are $P_1,\ldots,P_{r-1},P_{r+1},\ldots,P_n$ ( $r \in \{2,3,\ldots,n-1\}$ ).

On one hand, when $P_1,\ldots,P_{r-1},P_{r+1},\ldots,P_n$ launch active attacks on particles of $S_r^{2'}$ and $S_r^{2''}$, they actually act as an outside eavesdropper. As a result, their attacks inevitably leave trace on decoy photons so that they are detected by the eavesdropping check processes undoubtedly.

On the other hand, $P_1,\ldots,P_{r-1},P_{r+1},\ldots,P_n$ know the values of $-l_1^t, k_1^t \oplus l_1^t \oplus (-l_2^t), k_2^t \oplus l_2^t \oplus (-l_3^t), \ldots, k_{r-2}^t \oplus l_{r-2}^t \oplus (-l_{r-1}^t), k_r^t \oplus l_r^t \oplus (-l_{r+1}^t)$, where $t = 1,2,\ldots,N$, and can use $k_1^t, k_2^t, \ldots, k_{r-2}^t$ to decode out the value of $l_{r-1}^t$ from this information. When $P_1,\ldots,P_{r-1},P_{r+1},\ldots,P_n$ privately calculate Eq.(5) together with $P_r$, $P_1,\ldots,P_{r-1},P_{r+1},\ldots,P_n$ can know the value of $k_{r-1}^t \oplus l_{r-1}^t \oplus (-l_r^t)$, thus they can further decode out the value of $l_r^t$ according to $k_{r-1}^t$ and $l_{r-1}^t$. However, due to lack of the value of $l_{r+1}^t$, $P_1,\ldots,P_{r-1},P_{r+1},\ldots,P_n$ still cannot decode out $k_r^t$ from $k_r^t \oplus l_r^t \oplus (-l_{r+1}^t)$.

Finally, in Step 3, $P_0$ announces the comparison result of $K_1, K_2, \ldots, K_n$. However, $P_1,\ldots,P_{r-1},P_{r+1},\ldots,P_n$ cannot deduce $K_r$ just from the comparison result of $K_1, K_2, \ldots, K_n$.

③ The attack from TP

In order to obtain $k_1^t, k_2^t, \ldots, k_n^t$, $P_0$ may launch the $d$-level Bell entangled state measurement attack as follows: when $P_0$ and $P_i$ ( $i = 1,2,\ldots,n$ ) swap entanglement for particle sequences, $P_0$ performs $d$-level Bell entangled state measurement on particles $u_0^t$ and $v_i^t$ to obtain the value of $k_1^t \oplus l_1^t, k_2^t \oplus l_2^t, \ldots, k_n^t \oplus l_n^t$. Due to lack of the value of $l_1^t, l_2^t, \ldots, l_n^t$, $P_0$ cannot decode out $k_1^t, k_2^t, \ldots, k_n^t$ according to the value of $k_1^t \oplus l_1^t, k_2^t \oplus l_2^t, \ldots, k_n^t \oplus l_n^t$.

## 4 Discussion and conclusion

After ignoring the security check processes, the proposed scheme is compared with the MQPC schemes of Refs.[14-21] in detail. The comparison results are summarized in Table 1. Here, the qubit efficiency is defined as

$$\eta = \frac{c}{q}, \tag{7}$$

where $c$ and $q$ represent the number of compared classical bits and the number of consumed qubits, respectively. Since both the proposed scheme and the scheme of Ref.[19] use the quantum entanglement swapping within $d$-level quantum system to achieve the private comparison, we particularly illustrate the advantages the proposed scheme takes over the scheme of Ref.[19] here. It is easy to find out from Table 1 that, as long as $n$ is large enough, the qubit efficiency of the proposed scheme can be 1.5 times that of the scheme of Ref.[19]; and moreover, the proposed scheme exceeds the scheme of Ref.[19] in quantum resource and quantum measurement of TP, since the preparation and the quantum measurement of $d$-level two-particle Bell entangled states are easier than those of $d$-level $n+1$-particle Cat states.

Table 1  Comparison results between the proposed MQPC scheme and those of Refs.[14-21]

| | Quantum resource | Quantum measurement of TP | Quantum measurement of users | Unitary operation of TP | Unitary operation of users | Quantum memory | Qubit efficiency | Times of protocol execution | Quantum technology |
|---|---|---|---|---|---|---|---|---|---|
| Ref.[14] | n-particle GHZ class state | No | single-particle measurement | No | No | No | 1/n | 1 (whether the jth group bits from any two parties are equal or not equal) | The entanglement correlation among different particles from one quantum entangled state |
| Ref.[15] | d-level n-particle entangled states | d-level single-particle measurement | No | No | Yes | No | 1/n | 1 (whether the jth group bits from n parties are all equal or not all equal) | Unitary operation and quantum fourier transform |
| The first scheme of Ref.[16] | n-level n-particle entangled state and n-level two-particle entangled state | n-level single-particle measurement | n-level single-particle measurement | No | No | Yes | 1/3n | 1 (whether the jth group bits from n parties are all equal or not all equal) | Quantum fourier transform |
| The second scheme of Ref.[16] | n-level two-particle entangled state | n-level two-particle collective measurement | No | No | Yes | Yes | 1/2 (without considering the quantum resource consumed by a QKD scheme) | 1 (whether the jth group bits from n parties are all equal or not all equal) | Unitary operation |
| Ref.[17] | n-particle GHZ state | No | single-particle measurement | No | No | No | 1/n | 1 (whether the jth group bits from any two parties are equal or not equal) | The entanglement correlation among different particles from one quantum entangled state |
| Ref.[18] | Bell entangled state | Bell basis measurement | Bell basis measurement | No | No | Yes | 1/(n+1) | 1 (whether the jth group bits from any two parties are equal or not equal) | Quantum entanglement swapping |
| Ref.[19] | d-level n+1-particle Cat state and d-level two-particle Bell entangled state | d-level n+1-particle Cat state measurement | d-level two-particle Bell entangled state measurement | No | Yes | Yes | 1/(3n+1) | 1 (whether the jth group bits from n parties are all equal or not all equal) | Quantum entanglement swapping and unitary operation within d-level quantum system |
| Ref.[20] | Bell entangled state | single-particle measurement | single-particle measurement | No | No | Yes | 1/2n (without considering the quantum resource consumed by a QKD scheme) | 1 (whether the jth group bits from any two parties are equal or not equal) | The entanglement correlation among different particles from one quantum entangled state |
| Ref.[21] | n-level single-particle state | n-level single-particle measurement | No | No | Yes | No | 1/2 (without considering the quantum resource consumed by a QKD scheme) | 1 (whether the jth group bits from n parties are all equal or not all equal) | Unitary operation and quantum fourier transform |
| This paper | d-level two-particle Bell entangled state | d-level two-particle Bell entangled state measurement | d-level two-particle Bell entangled state measurement | No | Yes | Yes | 1/(2n+2) | 1 (whether the jth group bits from n parties are all equal or not all equal) | Quantum entanglement swapping and unitary operation within d-level quantum system |

To sum up, a MQPC scheme based on entanglement swapping of Bell entangled states within $d$-level quantum system is proposed in this paper, which can accomplish the equality comparison of secret binary sequences from $n$ users via one execution of scheme. Detailed security analysis shows that both the outside attack and the participant attack are ineffective. The proposed scheme needn't establish a private key among $n$ users beforehand through QKD method to encrypt the secret binary sequences. Compared with previous MQPC scheme based on $d$-level Cat states and $d$-level Bell entangled states, the proposed scheme has distinct advantages on quantum resource, quantum measurement of TP and qubit efficiency.

**Acknowledgement：**

Funding by the National Natural Science Foundation of China (Grant No.62071430) is gratefully acknowledged.

**References**

[1] Yao, A.C.: Protocols for secure computations. In: Proceedings of 23rd IEEE Symposium on Foundations of Computer Science (FOCS' 82), Washington, DC, USA, 1982, pp.160

[2] Boudot, F., Schoenmakers, B., Traor'e, J.: A fair and efficient solution to the socialist millionaires' problem. Discret Appl Math (Special Issue on Coding and Cryptology), 2001,111(1-2): 23-36

[3] Yang, Y.G., Wen, Q.Y.: An efficient two-party quantum private comparison protocol with decoy photons and two-photon entanglement. J Phys A : Math Theor, 2009, 42 : 055305

[4] Chen, X.B., Xu, G., Niu, X.X., Wen, Q.Y., Yang, Y.X.: An efficient protocol for the private comparison of equal information based on the triplet entangled state and single-particle measurement. Opt Commun, 2010,283:1561

[5] Yang, Y.G., Xia, J., Jia, X., Shi, L., Zhang, H.: New quantum private comparison protocol without entanglement. Int J Quantum Inf,2012, 10:1250065

[6] Tseng, H.Y., Lin, J., Hwang, T.: New quantum private comparison protocol using EPR pairs. Quantum Inf Process,2012,11:373-384

[7] Wang, C., Xu, G., Yang, Y.X.: Cryptanalysis and improvements for the quantum private comparison protocol using EPR pairs. Int J Quantum Inf,2013, 11:1350039

[8] Yang, Y.G., Xia, J., Jia, X., Zhang, H.: Comment on quantum private comparison protocols with a semi-honest third party. Quantum Inf Process, 2013, 12:877-885

[9] Zhang, W.W., Zhang, K.J.: Cryptanalysis and improvement of the quantum private comparison protocol with semi-honest third party. Quantum Inf Process, 2013, 12:1981-1990

[10] Chen, X.B., Su, Y., Niu, X.X., Yang, Y.X.: Efficient and feasible quantum private comparison of equality against the collective amplitude damping noise. Quantum Inf Process, 2014,13:101-112

[11] Ji, Z.X., Ye, T.Y.: Quantum private comparison of equal information based on highly entangled six-qubit genuine state. Commun Theor Phys, 2016, 65:711-715

[12] Ye, T.Y.: Quantum private comparison via cavity QED. Commun Theor Phys, 2017,67(2):147-156

[13] Ye, T.Y., Ji, Z.X.: Two-party quantum private comparison with five-qubit entangled states. Int J Theor Phys, 2017, 56(5):1517-1529

[14] Chang, Y.J., Tsai, C.W., Hwang, T.: Multi-user private comparison protocol using GHZ class states. Quantum Inf Process, 2013,12:1077-1088

[15] Liu, W., Wang, Y.B., Wang, X.M.: Multi-party quantum private comparison protocol using d-dimensional basis states without entanglement swapping. Int J Theor Phys, 2014, 53:1085-1091

[16] Wang, Q.L., Sun, H.X., Huang, W.: Multi-party quantum private comparison protocol with n-level entangled states. Quantum Inf Process, 2014, 13:2375-2389

[17] Hung, S.M., Hwang, S.L., Hwang, T., Kao, S.H.:Multiparty quantum private comparison with almost dishonest third parties for strangers. Quantum Inf Process, 2017,16(2):36

[18] Ye, T.Y.: Multi-party quantum private comparison protocol based on entanglement swapping of Bell entangled states. Commun Theor Phys, 2016, 66 (3):280-290

[19] Ji, Z.X., Ye, T.Y.: Multi-party quantum private comparison based on the entanglement swapping of d-level Cat states and d-level Bell states. Quantum Inf Process, 2017, 16(7): 177

[20] Ye, T.Y., Ji, Z.X.: Multi-user quantum private comparison with scattered preparation and one-way convergent transmission of quantum states. Sci China Phys, Mech and Astron, 2017, 60(9):090312

[21] Ye C.Q., Ye T.Y.: Circular multi-party quantum private comparison with n-level single-particle states. Int J Theor Phys, 2019,58(4):1282-1294

[22] Li, C. Y., Zhou, H.Y., Wang, Y., Deng, F.G.: Secure quantum key distribution network with Bell states and local unitary operations. Chin Phys Lett, 2005, 22(5):1049

[23] Li, C.Y., Li, X. H., Deng, F.G., Zhou, P., Liang, Y.J., Zhou, H.Y.: Efficient quantum cryptography network without entanglement and quantum memory. Chin Phys Lett, 2006, 23(11):2896

[24] Bennett, C.H., Brassard, G.: Quantum cryptography: public key distribution and coin tossing. In: Proceedings of the IEEE International Conference

on Computers, Systems and Signal Processing, Bangalore. 1984, pp.175-179
[25] Shor, P.W., Preskill, J.: Simple proof of security of the BB84 quantum key distribution protocol. Phys Rev Lett, 2000, 85(2):441
[26] Gao, F., Qin, S.J., Wen, Q.Y., Zhu, F.C.: A simple participant attack on the Bradler-Dusek protocol. Quantum Inf Comput, 2007, 7: 329